# Ferroelectric control of magnetic skyrmions in two-dimensional van der Waals heterostructures

Kai Huang,[1] Ding-Fu Shao,[2,*] and Evgeny Y. Tsymbal[1,†]

[1] Department of Physics and Astronomy & Nebraska Center for Materials and Nanoscience, University of Nebraska, Lincoln, Nebraska 68588-0299, USA

[2] Key Laboratory of Materials Physics, Institute of Solid State Physics, HFIPS, Chinese Academy of Sciences, Hefei 230031, China

**ABSTRACT:** Magnetic skyrmions are chiral nanoscale spin textures which are usually induced by Dzyaloshinskii–Moriya interaction (DMI). Recently, magnetic skyrmions have been observed in two-dimensional (2D) van der Waals (vdW) ferromagnetic materials, such as $Fe_3GeTe_2$. The electric control of skyrmions is important for their potential application in low-power memory technologies. Here, we predict that DMI and magnetic skyrmions in a $Fe_3GeTe_2$ monolayer can be controlled by ferroelectric polarization of an adjacent 2D vdW ferroelectric $In_2Se_3$. Based on density functional theory and atomistic spin-dynamics modeling, we find that the interfacial symmetry breaking produces a sizable DMI in a $Fe_3GeTe_2/In_2Se_3$ vdW heterostructure. We show that the magnitude of DMI can be controlled by ferroelectric polarization reversal, leading to creation and annihilation of skyrmions. Furthermore, we find that the sign of DMI in a $In_2Se_3/Fe_3GeTe_2/In_2Se_3$ heterostructure changes with ferroelectric switching reversing the skyrmion chirality. The predicted electrically controlled skyrmion formation may be interesting for spintronic applications.

Magnetic skyrmions are topological magnetic quasiparticles exhibiting a whirling spin texture in real space [1]. These spin-chiral objects have recently attracted significant interest due to the rich physics and promising spintronic applications [2-6]. The formation of magnetic skyrmions usually requires a strong Dzyaloshinskii-Moriya interaction (DMI) – an exchange interaction between adjacent magnetic moments driven by structural asymmetry and spin-orbit coupling [7-9]. Bulk magnets hosting magnetic skyrmions are limited to those with chiral crystal structures that support a finite DMI [2,10-13]. The skyrmions in these chiral magnets are usually observed at low temperature, which limits potential applications. Alternatively, a strong DMI can be induced by symmetry breaking and strong spin-orbit coupling at interfaces in the multilayer films composed of magnetic and heavy-metal layers [2,14-16]. This allows the formation of skyrmions at high temperature.

Currently, the efforts are aimed at exploring the material systems producing stable magnetic skyrmions that can be conveniently manipulated by external stimulus. Typically, the generation of magnetic skyrmions requires an external magnetic field to adjust a delicate balance between the exchange, anisotropy, DMI, and Zeeman energy contributions controlling different types of spin textures. Using an electric field is however a more energy-efficient method to control skyrmions which can be used in low-power spintronics. It has been demonstrated that magnetic skyrmions can be created and annihilated by voltage-controlled exchange coupling or magnetic anisotropy [17,18]. Another approach to electrically control skyrmions is to exploit switchable polarization of an adjacent ferroelectric material [19,20]. This method has been shown to efficiently manipulate skyrmions in magnetic thin films interfaced with perovskite ferroelectrics [21-24].

The recent discoveries of two-dimensional (2D) van der Waals (vdW) materials exhibiting spontaneous electric or magnetic polarizations [25-38] opened a new dimension in exploring and exploiting ferroic and topological properties of materials including magnetic skyrmions. In a 2D ferromagnet, the symmetry breaking required for DMI to be realized by the interfacial proximity effect in a vdW heterostructure [39]. This allows the stabilization of magnetic skyrmions in a device with minimum thickness and even in the absence of the external magnetic field, as has been demonstrated in the recent experiments [40-43].

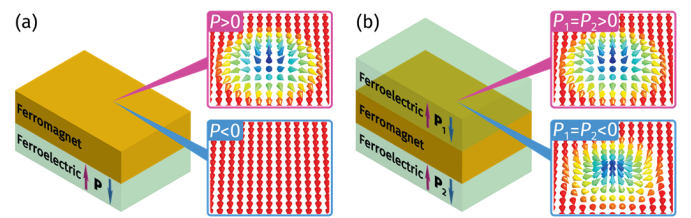

**FIG. 1**: (**a**) Schematic of a vdW heterostructure composed of a 2D ferromagnet and a 2D ferroelectric, where the switching of electric polarization ($\vec{P}$) can reversibly create and annihilate magnetic skyrmions. (**b**) Schematic of a vdW heterostructure where a 2D ferromagnet is sandwiched between two identical 2D ferroelectrics. The simultaneous switching of the polarizations of the top and bottom ferroelectrics ($\vec{P}_1$ and $\vec{P}_2$) changes the chirality of magnetic skyrmions.

A nonvolatile control of magnetic skyrmions by electric field is desirable for device applications. The recently discovered 2D ferroelectric materials [30-31,44-45], such as In$_2$Se$_3$ [46-49], can be used to mediate this effect. Due to their switchable electric polarization [50-54], the 2D ferroelectrics can provide a voltage tunable interface proximity effect to other 2D materials. In particular, the electric polarization of a 2D ferroelectric is expected to efficiently control DMI in an adjacent 2D ferromagnet. This is due to DMI being sensitive to the interface orbital hybridizations and charge transfer effects [55]. As a result, when heterostructured with 2D ferromagnets, 2D ferroelectrics can be used to mediate the electrical control DMI and thus skyrmion behaviors [56,57].

Fe$_3$GeTe$_2$ is a representative 2D ferromagnetic metal [58] which can be used as a viable material to explore the control of skyrmions by ferroelectric polarization. Experimentally, the emergence of skyrmions has been demonstrated in Fe$_3$GeTe$_2$, under certain conditions [40-43]. Theoretically, when interfaced with a 2D ferroelectric In$_2$Se$_3$ layer (Fig. 1(a)), different magnitudes of DMI are expected in Fe$_3$GeTe$_2$ depending ferroelectric polarization of In$_2$Se$_3$. This may lead to the reversable creation and annihilation of magnetic skyrmions. A Fe$_3$GeTe$_2$ monolayer can be further sandwiched between two identical ferroelectric In$_2$Se$_3$ layers (Fig. 1(b)). In this case, a simultaneous switching of polarization in both layers is equivalent to applying a symmetry operation to the whole system, which is expected to change the sign of DMI and thus reverse the chirality of magnetic skyrmions.

In this work, using first-principles density functional theory (DFT) calculations, we demonstrate that DMI in a Fe$_3$GeTe$_2$ monolayer can be induced and controlled when it is interfaced with a 2D ferroelectric In$_2$Se$_3$ layer. In such a Fe$_3$GeTe$_2$/In$_2$Se$_3$ vdW heterostructure, we find that the reversal of ferroelectric polarization of In$_2$Se$_3$ switches the DMI from a large to small value. As a result, based on our atomistic spin-dynamics modeling, we predict the reversable creation and annihilation of skyrmions in the system. Further, we show that the DMI sign in a In$_2$Se$_3$/Fe$_3$GeTe$_2$/In$_2$Se$_3$ sandwich structure changes with ferroelectric polarization switching, resulting in the reversal of magnetic skyrmion chirality. Our results show a possibility of the nonvolatile control of magnetic skyrmions in monolayer Fe$_3$GeTe$_2$ by an electric field, which may be interesting for the potential application of skyrmion systems.

Fe$_3$GeTe$_2$ is a hexagonal layered ferromagnetic metal with the easy axis along the [001] direction [59-61]. Each Fe$_3$GeTe$_2$ layer contains five atomic layers stacked with Te-Fe$_I$-GeFe$_{II}$-Fe$_I$-Te sequence, where Fe$_I$ denotes the upmost and downmost Fe atoms, and Fe$_{II}$ denotes the Fe atom in the central GeFe plane (Fig. 2(a,b)). In its bulk phase, adjacent Fe$_3$GeTe$_2$ monolayers are rotated by 180° with respect to each other, forming a centrosymmetric structure with a magnetic point group $6/mm'm'$. The presence of inversion symmetry in bulk Fe$_3$GeTe$_2$ prohibits a finite DMI and hence magnetic skyrmions are not expected to emerge in bulk Fe$_3$GeTe$_2$.

Recently, Fe$_3$GeTe$_2$ has been successfully exfoliated down to a monolayer [58,59]. In the absence of the bulk interlayer stacking, monolayer Fe$_3$GeTe$_2$ is noncentrosymmetric and belongs to the magnetic point group $\bar{6}m'2'$. Although the inversion symmetry is broken, the DMI in monolayer Fe$_3$GeTe$_2$ is still vanishing due to other symmetry operations [62], such as (001) mirror plane reflection $\hat{M}_\perp$ (Fig. 2(a)), and symmetry operation $\hat{T}\hat{C}_{2\parallel}$ combining time reversal symmetry $\hat{T}$ and two-fold rotation symmetry $\hat{C}_{2\parallel}$ with respect to the three in-plane axes along the in-plane directions of the nearest Fe neighbors (Fig. 2(b)). The absence of DMI in monolayer Fe$_3$GeTe$_2$ can be understood by comparing the energies of two artificial magnetic configurations with opposite chirality. As schematically shown in Fig. 2(c), the clockwise (CW) and counterclockwise (CCW) configurations of Fe$_3$GeTe$_2$ can be transformed to each other by a $\hat{M}_\perp$ operation. This is due to $\hat{M}_\perp$ reversing the in-plane components of the Fe moments but conserving the out-of-plane component, i.e.

$$\hat{M}_\perp(m_x, m_y, m_z) = (-m_x, -m_y, m_z). \quad (1)$$

As a result, the DMI energy $E_{\text{DMI}}$ which is determined by the energy difference between the CW and CCW states [63]

$$E_{\text{DMI}} \propto E_{\text{CW}} - E_{\text{CCW}}, \quad (2)$$

is zero due to the equal values of $E_{\text{CW}}$ and $E_{\text{CCW}}$ enforced by $\hat{M}_\perp$. Similarly, $\hat{T}\hat{C}_{2\parallel}$ forbids the chiral magnetic configurations and results in zero $E_{\text{DMI}}$ in monolayer Fe$_3$GeTe$_2$.

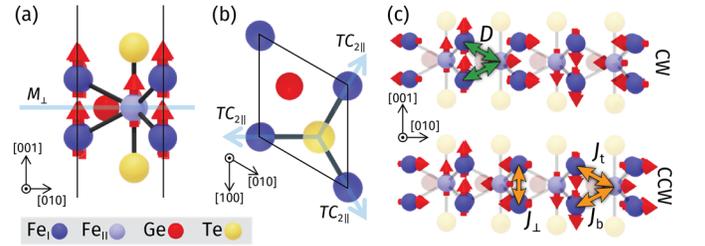

**FIG. 2**: (**a,b**) The atomic structure of monolayer Fe$_3$GeTe$_2$ – side view (**a**) and top view (**b**). (**c**) The artificial magnetic configurations of clockwise (CW) and counterclockwise (CCW) chirality along the [010] direction. Red arrows denote the magnetic moment of Fe. The horizontal blue line in (**a**) denotes the mirror plane $\hat{M}_\perp$. Blue arrows in (**b**) denote the symmetry operations $\hat{T}\hat{C}_{2\parallel}$ combining time reversal symmetry $\hat{T}$ and two-fold rotation symmetry $\hat{C}_{2\parallel}$ with respect to the three in-plane axes. Orange and green arrows in (**c**) denote the Heisenberg exchange parameters ($J$) and DMI coefficient ($D$) between the Fe moments, respectively.

These symmetry constraints can be broken if the top and bottom surfaces of the monolayer Fe$_3$GeTe$_2$ become asymmetric due to an interface proximity effect. Here, we consider the proximity of 2D ferroelectric In$_2$Se$_3$ interfaced with monolayer Fe$_3$GeTe$_2$. A monolayer In$_2$Se$_3$ contains five triangular lattices stacked with Se-In-Se-In-Se sequence. The central Se atom is located at one of the two asymmetric but topologically identical sites, associated with a finite out-of-plane polarization $\vec{P}$ pointing in opposite directions [34,49]. Recent reports show that In$_2$Se$_3$ can be effectively used in vdW heterostructures to provide a nonvolatile control of 2D electronic structures despite of very weak interlayer coupling [50-53]. Due to both having hexagonal atomic structures and similar in-plane lattice constants [60,64], Fe$_3$GeTe$_2$ and In$_2$Se$_3$ are well matched to construct commensurate vdW heterostructures. In such heterostructures, the $\hat{M}_\perp$ and $\hat{T}\hat{C}_{2\parallel}$ symmetries are broken, making DMI finite and thus the appearance of magnetic skyrmions possible. The DMI can be further controlled by ferroelectric polarization of In$_2$Se$_3$, resulting in variable behaviors of magnetic skyrmions.

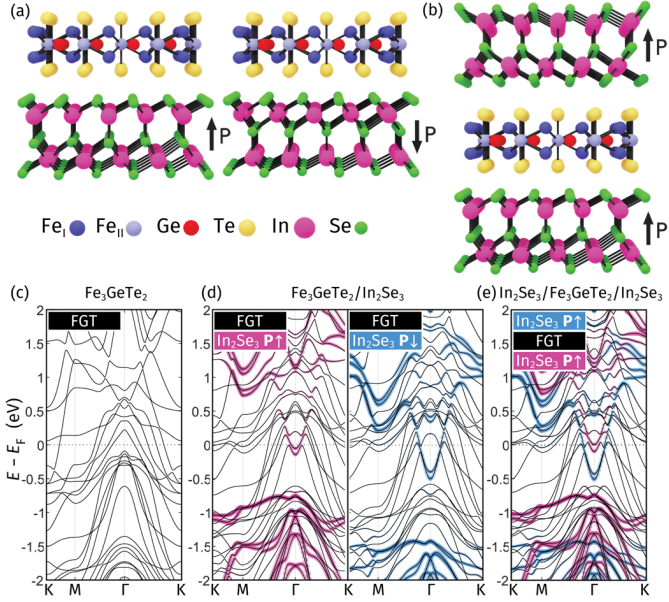

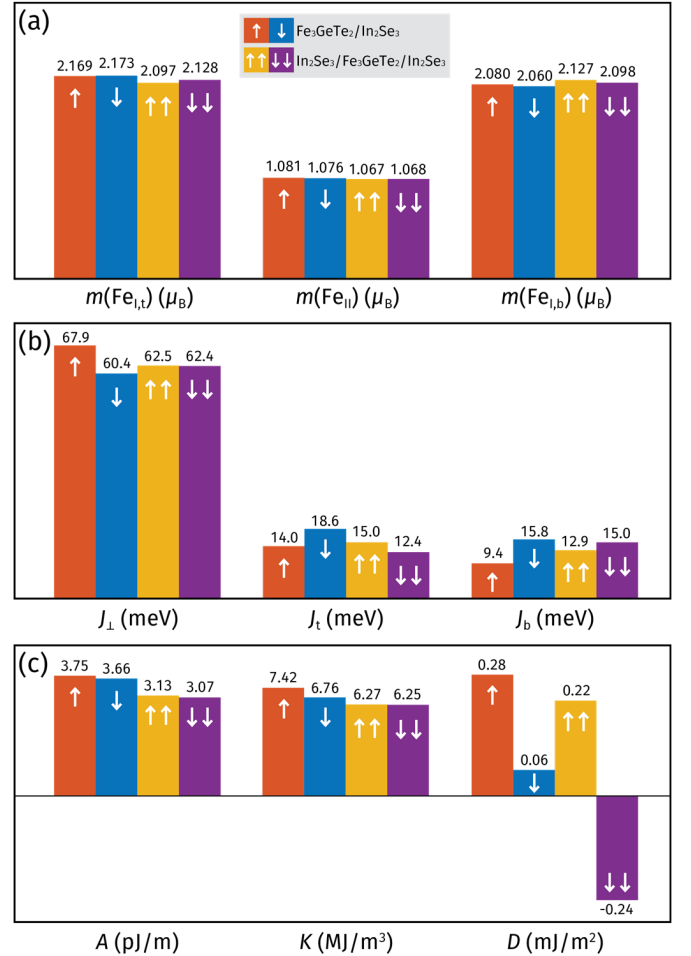

FIG. 3: (**a**) Fe$_3$GeTe$_2$/In$_2$Se$_3$ vdW heterostructure for opposite polarizations in In$_2$Se$_3$. (**b**) In$_2$Se$_3$/Fe$_3$GeTe$_2$/In$_2$Se$_3$ vdW heterostructure. (**c**) The band structure of a freestanding Fe$_3$GeTe$_2$ monolayer. (**d**) The band structure of Fe$_3$GeTe$_2$/In$_2$Se$_3$ for In$_2$Se$_3$ polarization pointing up (left) and down (right). (**e**) The band structure of In$_2$Se$_3$/Fe$_3$GeTe$_2$/In$_2$Se$_3$. The colored dots in (**d,e**) indicate the bands contributed by In$_2$Se$_3$.

First, we ascertain that our calculations correctly predict the ground state of monolayer Fe$_3$GeTe$_2$ and monolayer In$_2$Se$_3$. For Fe$_3$GeTe$_2$, the magnetic moments of the Fe atoms are calculated to be $m(\text{Fe}_\text{I}) = 2.18\ \mu_B$ and $m(\text{Fe}_\text{II}) = 1.09\ \mu_B$. The total energy calculation predicts an out-of-plane anisotropy with the magnetocrystalline anisotropy energy $K = E_{[100]} - E_{[010]} = 3.07$ meV/f. u. $= 8.57$ MJ/m$^3$. These values are consistent well with the previous calculations [61]. The band gap and ferroelectric polarization of In$_2$Se$_3$ are estimated to be 0.64 eV and 1.45 µC/cm$^2$, respectively, slightly different to these reported in Ref. [49] but do not influence the results in this work (see supporting information).

We then construct a vdW heterostructure by attaching monolayer Fe$_3$GeTe$_2$ on top of monolayer In$_2$Se$_3$ (Fig. 3(a)). To ensure that the Fe$_3$GeTe$_2$/In$_2$Se$_3$ stacking corresponds to the ground state, we calculate the total energy of the bilayer as a function of the lateral alignment of the two monolayers (see Supplemental Material for details). We find that the ground state corresponds to the stacking order where Fe$_\text{I}$ atoms lie atop the nearest In atom (Fig. 3(a)).

Figures 3 (c,d) show the band structure of the freestanding monolayer Fe$_3$GeTe$_2$ in comparison to that of the bilayer Fe$_3$GeTe$_2$/In$_2$Se$_3$. We find that the Fe$_3$GeTe$_2$ electronic bands in the bilayer structure (Fig. 3(d)) do not change much compared to those in the freestanding Fe$_3$GeTe$_2$ (Fig. 3(c)). This is due to negligible orbital hybridization across the interface reflecting weak vdW interactions. The bilayer bands exhibit a rigid shift toward higher energy resulting from a different chemical potential in the heterostructure. The electronic bands contributed by In$_2$Se$_3$ in the bilayer (denoted by colored dots) are also similar to those in its freestanding form (not shown). However, the conduction band minimum of In$_2$Se$_3$ is shifted below the Fermi level ($E_F$) (Figs. 3(d)), due to the electron charge transfer from Fe$_3$GeTe$_2$ to In$_2$Se$_3$. Polarization switching from up to down does not notably change the bands contributed by Fe$_3$GeTe$_2$. However, it shifts the In$_2$Se$_3$ bands to lower energy due to the field effect resulting from the polarization charge of In$_2$Se$_3$.

FIG. 4: Calculated magnetic parameters for Fe$_3$GeTe$_2$/In$_2$Se$_3$ and In$_2$Se$_3$/Fe$_3$GeTe$_2$/In$_2$Se$_3$ heterostructures. (**a**) Magnetic moments of Fe atoms. (**b**) Heisenberg exchange parameters ($J$) denoted in Fig. 2(c). (**c**) Exchange stiffness ($A$), magnetic anisotropy ($K$), and the DMI coefficient ($D$). The single arrows (↑ or ↓) and double arrows (↑↑ or ↓↓) denote the polarization direction in Fe$_3$GeTe$_2$/In$_2$Se$_3$ and In$_2$Se$_3$/Fe$_3$GeTe$_2$/In$_2$Se$_3$, respectively.

Figure 4 shows results of our calculations of the magnetic parameters for the Fe$_3$GeTe$_2$/In$_2$Se$_3$ bilayer. We find that the magnetic moments of the top Fe$_\text{I}$ atom (denoted by Fe$_{\text{I,t}}$) and the central Fe$_\text{II}$ atom do not change much compared to those in the freestanding Fe$_3$GeTe$_2$. On the contrary, the magnetic moment of the bottom Fe$_\text{I}$ atom (denoted by Fe$_{\text{I,b}}$), which lies closer to the interface with In$_2$Se$_3$, is reduced by about ∼0.1 $\mu_B$ compared to that in the freestanding Fe$_3$GeTe$_2$. This is due to the electron charge transfer from Fe$_3$GeTe$_2$ to In$_2$Se$_3$ not related to ferroelectric polarization. The polarization switching in In$_2$Se$_3$ does not produce notable changes in the magnetic moments due to the small polarization of In$_2$Se$_3$. This is consistent with the similar band structures contributed by $Fe_3GeTe_2$ in the $Fe_3GeTe_2/In_2Se_3$ bilayer for different polarizations.

The difference in the magnetic moments of the $Fe_{I,t}$ and $Fe_{I,b}$ atoms reflects the broken mirror symmetry in the bilayer. As a result, the Heisenberg exchange coupling between $Fe_{I,t}$ and $Fe_{II}$ ($J_t$) is different from that between $Fe_{I,b}$ and $Fe_{II}$ ($J_b$) (Fig. 4(b)). We find that the changes of the Heisenberg exchange parameters and the derived exchange stiffness ($A$) induced by ferroelectric switching are not significant (Fig. 4(b,c)) due to the small ferroelectric polarization in $In_2Se_3$. Similarly, change in the magnetic anisotropy parameter ($K$) in response to ferroelectric switching is also not large (Fig. 4(c)).

The DMI coefficient $D$ can be calculated using [63]

$$D = \frac{E_{CW} - E_{CCW}}{4\sqrt{3}ah}, \quad (3)$$

where $a, h$ are the lattice constant and layer thickness, respectively. The finite $D$ is supported by the magnetic point group $3m'1$ of the $Fe_3GeTe_2/In_2Se_3$ heterostructure due to broken $\hat{M}_\perp$ and $\hat{T}\hat{C}_{2\parallel}$ symmetries. We find a large $D_\uparrow = 0.28$ mJ/m$^2$ (DMI for polarization of the $In_2Se_3$ is pointing upward) and a small $D_\downarrow = 0.06$ mJ/m$^2$ (DMI for polarization of the $In_2Se_3$ is pointing downward). Such a sizable change in the DMI induced by polarization switching is in contrast to the relatively small changes in other magnetic properties. This is due to DMI being very sensitive to the structural asymmetry, which can be effectively controlled by ferroelectric switching.

Next, we construct a $In_2Se_3/Fe_3GeTe_2/In_2Se_3$ vdW heterostructure by adding an additional monolayer $In_2Se_3$ on the top of $Fe_3GeTe_2/In_2Se_3$ (Fig. 3(b)). In order to ensure the low energy stacking at both interfaces, the top $In_2Se_3$ layer is obtained by applying a mirror reflection of the bottom $In_2Se_3$ monolayer. Figure 3(e) shows the calculated band structure of the $In_2Se_3/Fe_3GeTe_2/In_2Se_3$ trilayer with polarizations pointing upward for both $In_2Se_3$ layers. We find that the electronic bands contributed by $Fe_3GeTe_2$ show negligible changes compared to those in the $Fe_3GeTe_2/In_2Se_3$ bilayer. At the same time, the bands contributed by $In_2Se_3$ can be considered as a superposition of these for $Fe_3GeTe_2/In_2Se_3$ with opposite polarizations (Figs. 3(c,d)).

Figure 4 shows the calculated magnetic parameters for the $In_2Se_3/Fe_3GeTe_2/In_2Se_3$ heterostructure. The magnitudes of the magnetic moments, Heisenberg exchange parameters, and magnetic anisotropy reveal slight changes compared to these of the $Fe_3GeTe_2/In_2Se_3$ structure. As expected, the magnetic moments $m(Fe_{I,t})$ and $m(Fe_{I,b})$ as well as the exchange constants $J_t$ and $J_b$ swap their values upon the simultaneous switching of polarizations of both $In_2Se_3$ layers. This is because such a switching is equivalent to applying $\hat{T}\hat{C}_{2\parallel}$ operation to the system, which swap the moments and exchange parameters on top and bottom of the $GeFe_{II}$ plane.

The DMI coefficient for polarization of both $In_2Se_3$ layers pointing upward (downward) is calculated to be $D_{\uparrow\uparrow} \approx 0.22$ mJ/m$^2$ ($D_{\downarrow\downarrow} \approx -0.24$ mJ/m$^2$). We see that within the calculation accuracy, $D_{\uparrow\uparrow} \approx -D_{\downarrow\downarrow}$, i.e. the sign of the DMI coefficient changes with polarization switching, which is due to ferroelectric switching being equivalent to the $\hat{T}\hat{C}_{2\parallel}$ symmetry transformation. Comparing the DMI coefficients for the $Fe_3GeTe_2/In_2Se_3$ and $In_2Se_3/Fe_3GeTe_2/In_2Se_3$ structures (Fig. 4), we find $D_{\uparrow\uparrow} \approx D_\uparrow - D_\downarrow$. This result follows from the presence of two interfaces (top and bottom) in the $In_2Se_3/Fe_3GeTe_2/In_2Se_3$ trilayer, where at one interface the polarization is pointing toward the $Fe_3GeTe_2$ layer and at the other interface the polarization is pointing away from the $Fe_3GeTe_2$ layer. Therefore, the interfacial proximity effect on the DMI constant in the $In_2Se_3/Fe_3GeTe_2/In_2Se_3$ trilayer is a sum of the two contributions from two $Fe_3GeTe_2/In_2Se_3$ interfaces.

The predicted ferroelectric switching of DMI in the $Fe_3GeTe_2/In_2Se_3$ and $In_2Se_3/Fe_3GeTe_2/In_2Se_3$ heterostructures indicates a possibility of the electric field control of magnetic skyrmions in these systems. To demonstrate this effect, we perform the atomistic spin-dynamics modeling using LLG equation

$$\frac{\partial \vec{m}_i}{\partial t} = -\gamma(\vec{m}_i \times \vec{H}_i) + \alpha_G \left(\vec{m}_i \times \frac{\partial \vec{m}_i}{\partial t}\right). \quad (4)$$

Here $\alpha_G$ is the Gilbert damping constant, $\gamma$ is the gyromagnetic ratio, $\vec{m}_i = \frac{\vec{M}_i}{|\vec{M}_i|}$ is the unit magnetization vector for each sublattice with the magnetization $\vec{M}_i$. The magnetic field $\vec{H}_i = -\frac{1}{\mu}\frac{\partial H}{\partial \vec{m}_i}$ is determined by the spin Hamiltonian:

$$H = -\sum_{i \neq j} J_{ij}\vec{m}_i \cdot \vec{m}_j + \sum_{i \neq j} \vec{d}_{ij} \cdot (\vec{m}_i \times \vec{m}_j) - K\sum_i (\hat{n}_i \cdot \vec{m}_i)^2 \quad (5)$$

where $\mu$ is the magnetic moment of a Fe atom, $J_{ij}$ is the exchange coupling, $\vec{d}_{ij}$ is the DMI vector, $K$ is the magnetic anisotropy energy per Fe atom, and $\hat{n}_i$ is the direction of the easy axis.

A supercell of $Fe_3GeTe_2$ monolayer with the size of 60 nm × 60 nm is used in our simulation. In this supercell, nonmagnetic Ge and Te atoms are ignored, only Fe honeycomb lattice is kept. A round ferromagnetic domain is initially set in the center of the supercell, where the magnetic moments are pointing along +z/-z directions inside/outside the domain wall. The magnetic configuration of the system is then relaxed to the equilibrium state. The round domain can gradually relax to a stable skyrmion or shrink and eventually disappear in the background depending on the magnetic parameters used in the simulation.

We find no magnetic skyrmion emerging in our atomistic modeling with the magnetic parameters given in Fig. 4. This can be explained from the expected radius ($R$) of a skyrmion [65]

$$R = \pi D \sqrt{\frac{A}{16AK^2 - \pi^2 D^2 K}}. \quad (6)$$

For the calculated magnetic parameters in Fig. 4, we obtain the largest $R$ to be $\sim 0.03$ nm, which is too small to observe. This is due to a very large magnetic anisotropy $K$ predicted theoretically for $Fe_3GeTe_2$ [61] compared to the recent experimental measurements where magnetic skyrmions were observed in various $Fe_3GeTe_2$ based systems [40-43].

It is known that the magnetic anisotropy of $Fe_3GeTe_2$ can be strongly suppressed by many factors such as doping and temperature [66-68]. Indeed, if we use the calculated $A$ and $D$ parameters and the radius of the magnetic skyrmions $R \sim 100$ nm reported experimentally [40-43], a much weaker magnetic anisotropy $K \sim 0.008$ MJ/m$^3$ is estimated by Eq. (4). We thus assume a moderate anisotropy value of $K \sim 0.04$ MJ/m$^3$ for our modelling to qualitatively demonstrate ferroelectric effect on the appearance of skyrmions in the considered systems.

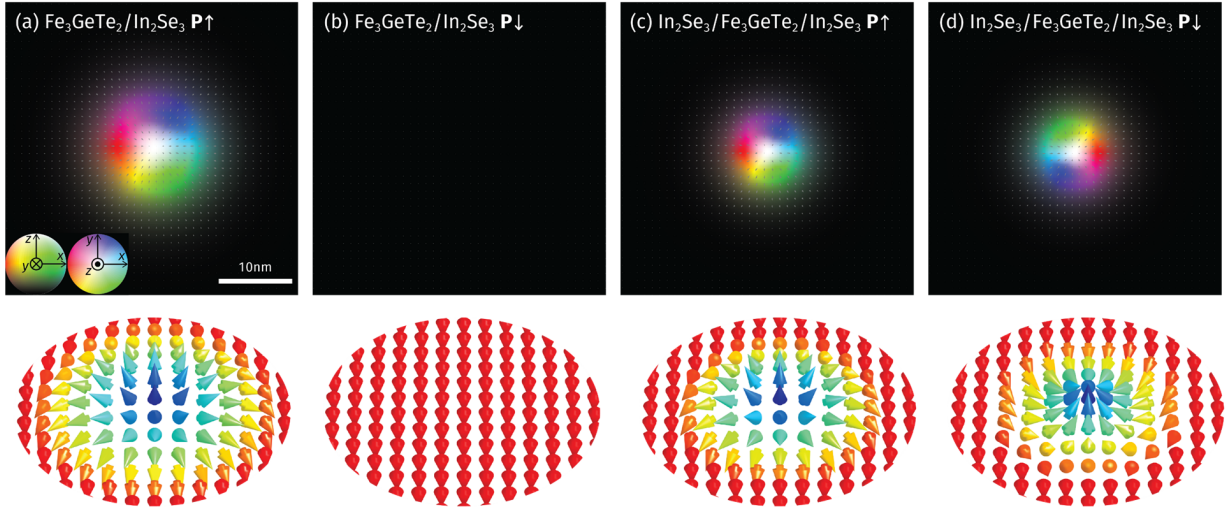

**FIG. 5**: (**a,b,c,d**) Results of atomistic spin-dynamics modeling of the Fe$_3$GeTe$_2$/In$_2$Se$_3$ vdW heterostructure (**a,b**) and the In$_2$Se$_3$/Fe$_3$GeTe$_2$/In$_2$Se$_3$ vdW heterostructure (**c,d**) for different ferroelectric polarizations.

With this value of $K$, the results of our atomistic spin-dynamics modeling predict for the Fe$_3$GeTe$_2$/In$_2$Se$_3$ heterostructure, that a Néel-type skyrmion of about 12 nm in diameter emerges when the polarization is pointing upward (Fig. 5(a)) and disappears when the polarization is pointing downward (Fig. 5(b)). Such creation and annihilation of a magnetic skyrmion in this heterostructure is due to the change of DMI with reversal of ferroelectric polarization of In$_2$Se$_3$. For the In$_2$Se$_3$/Fe$_3$GeTe$_2$/In$_2$Se$_3$ heterostructure, we obtain a skyrmion of about 6 nm in diameter (Fig. 5(c)). As expected, its chirality is reversed with reversal of polarization of In$_2$Se$_3$ (Fig. 5(d)) due to the change of the DMI sign.

Previous investigations showed that magnetic skyrmions can be moved at a high speed by a moderate electric current, indicating promising spintronic applications [1]. A well-controlled writing and erasing process is required prior to such applications. The creation and annihilation of skyrmions by voltage is desirable for this purpose, due to low energy consumption. Our work demonstrates that the required functional behavior can be achieved in a Fe$_3$GeTe$_2$/In$_2$Se$_3$ system where the interfacial proximity of a 2D magnet and a 2D ferroelectric induces DMI dependent on ferroelectric polarization. The controllable chirality of skyrmions in a In$_2$Se$_3$/Fe$_3$GeTe$_2$/In$_2$Se$_3$ heterostructure may be interesting for programmable skyrmion-based memories and logic [69]. The proposed approach is not limited to the Fe$_3$GeTe$_2$/In$_2$Se$_3$ system, but should be valid for vdW heterostructures constructed from other 2D ferroics.

In conclusion, based on first-principles DFT calculations and atomistic spin-dynamics modelling, we have demonstrated that the Dzyaloshinskii–Moriya interaction in monolayer Fe$_3$GeTe$_2$ can be induced and controlled by ferroelectric polarization of an adjacent ferroelectric In$_2$Se$_3$. For a Fe$_3$GeTe$_2$/In$_2$Se$_3$ vdW heterostructure, we have predicted that the reversal of ferroelectric polarization of In$_2$Se$_3$ can switch the DMI of the system from a large to small value, resulting in the reversible skyrmion creation and annihilation. For a In$_2$Se$_3$/Fe$_3$GeTe$_2$/In$_2$Se$_3$ vdW heterostructure, we have shown that the DMI sign was changed with ferroelectric switching, resulting in the reversal of chirality of the magnetic skyrmions. Our work demonstrated a possibility of the nonvolatile control of magnetic skyrmions in monolayer Fe$_3$GeTe$_2$ by an applied electric field, which is promising for the potential device application of skyrmion systems.

## ASSOCIATED CONTENT

**Supporting Information**

Additional details of computational methods; theoretical derivation of the DMI vector; calculation of the DMI vectors in Fe$_3$GeTe$_2$ based systems; total energy calculations of the most stable Fe$_3$GeTe$_2$/In$_2$Se$_3$ stacking order.

## AUTHOR INFORMATION

**Corresponding Authors**

Ding-Fu Shao – https://orcid.org/0000-0002-2732-4131; Email: dfshao@issp.ac.cn
Evgeny Y. Tsymbal – https://orcid.org/0000-0002-6728-5480; Email: tsymbal@unl.edu

**Authors**

Kai Huang – https://orcid.org/0000-0001-5527-3426; Email: kai.huang@huskers.unl.edu**Notes**

The authors declare no competing financial interest.

## ACKNOWLEDGMENTS

The authors thank Robert Streubel for stimulating discussions. This work was supported by the by the EPSCoR RII Track-1 (NSF Award OIA-2044049) program. Computations were performed at the University of Nebraska Holland Computing Center.

(1) Rößler, U. K.; Bogdanov, A. N.; Pfleiderer, C. Spontaneous skyrmion ground states in magnetic metals. *Nature* **2006**, 442, 797–801.

(2) Fert, A.; Reyren, N.; Cros, V. Magnetic skyrmions: advances in physics and potential applications. *Nat. Rev. Mater.* **2017**, 2, 17031.

(3) Tokura, Y.; Tokura, Y. Magnetic skyrmion materials. *Chem. Rev.* **2021**, 121, 2857–2897.

(4) Göbel, B.; Mertig, I.; Tretiakov, O. A. Beyond skyrmions: Review and perspectives of alternative magnetic quasiparticles. *Phys. Rep.* **2021**, 895, 1–28.

(5) Everschor-Sitte, K.; Masell, J.; Reeve, R. M.; Kläui, M. Perspective: Magnetic skyrmions—Overview of recent progress in an active research field. *J. Appl. Phys.* **2018**, 124, 240901.

(6) Nagaosa, N.; Tokura, Y. Topological properties and dynamics of magnetic skyrmions. *Nature Nanotech* **2013**, 8, 899–911.

(7) Dzyaloshinsky, I. A thermodynamic theory of "weak" ferromagnetism of antiferromagnetics. *J. Phys. Chem. Solids* **1958**, 4, 4.

(8) Moriya, T. New Mechanism of Anisotropic Superexchange Interaction. *Phys. Rev. Lett.* **1960**, 4, 228.

(9) Moriya, T. Anisotropic superexchange interaction and weak ferromagnetism. *Phys. Rev.* **1960**, 120, 91.

(10) Mühlbauer, S.; Binz, B.; Jonietz, F.; Pfleiderer, C.; Rosch, A.; Neubauer, A.; Georgii, R.; Böni, P. Skyrmion Lattice in a Chiral Magnet. *Science* **2009**, 323, 5916.

(11) Tanigaki, T.; Shibata, K.; Kanazawa, N.; Yu, X.; Onose, Y.; Park, H. S.; Shindo, D.; Tokura, Y. Real-space observation of short-period cubic lattice of skyrmions in MnGe. *Nano Lett.* **2015**, 15, 5438–5442.

(12) Seki, S.; Yu, X. Z.; Ishiwata, S.; Tokura, Y. Observation of Skyrmions in a Multiferroic Material. *Science* **2012**, 336, 198–201.

(13) Kézsmárki, I.; Bordács, S.; Milde, P.; Neuber, E.; Eng, L. M.; White, J. S.; Rønnow, H. M.; Dewhurst, C. D.; Mochizuki, M.; Yanai, K.; Nakamura, H.; Ehlers, D.; Tsurkan, V.; Loidl, A. Néel-type skyrmion lattice with confined orientation in the polar magnetic semiconductor $GaV_4S_8$. *Nature Mater.* **2015**, 4, 1116–1122.

(14) Heinze, S.; von Bergmann, K.; Menzel, M.; Brede, J.; Kubetzka, A.; Wiesendanger, R.; Bihlmayer, G.; Blügel, S. Spontaneous atomic-scale magnetic skyrmion lattice in two dimensions. *Nature Phys.* **2011**, 7, 713–718.

(15) Boulle, O.; Vogel, J.; Yang, H.; Pizzini, S.; Chaves, D. de Souza; Locatelli, A.; Menteş, T. O.; Sala, A.; Buda-Prejbeanu, L. D.; Klein, O.; Belmeg-uenai, M.; Roussigné, Y.; Stashkevich, A.; Chérif, S. M.; Aballe, L.; Foerster, M.; Chshiev, M.; Auffret, S.; Miron, I. M.; Gaudin, G. Room-temperature chiral magnetic skyrmions in ultrathin magnetic nanostructures. *Nature Nanotech.* **2016**, 11, 449–454.

(16) Woo, S.; Litzius, K.; Krüger, B.; Im, M.-Y.; Caretta, L.; Richter, K.; Mann, M.; Krone, A.; Reeve, R. M.; Weig, M.; Agrawal, P.; Lemesh, I.; Mawass, M.-A.; Fischer, P.; Kläui, M.; Beach, G. S. D. Observation of room-temperature magnetic skyrmions and their current-driven dynamics in ultrathin metallic ferromagnets. *Nature Mater.* **2016**, 15, 501–506.

(17) Hsu, P.-J.; Kubetzka, A.; Finco, A.; Romming, N.; von Bergmann, K.; Wiesendanger R. Electric-field-driven switching of individual magnetic skyrmions. *Nature Nanotech.* **2017**, 12, 123–126.

(18) Bhattacharya, D.; Razavi, S. A.; Wu, H.; Dai, B.; Wang, K. L.; Atulasimha, J. Creation and annihilation of non-volatile fixed magnetic skyrmions using voltage control of magnetic anisotropy. *Nature Electronics* **2020**, 3, 539–545.

(19) Liang, J.; Cui, Q.; Yang, H. Electrically switchable Rashba-type Dzyaloshinskii-Moriya interaction and skyrmion in two-dimensional magnetoelectric multiferroics. *Phys. Rev. B* **2020**, 102, 220409(R).

(20) Cui, Q.; Zhu, Yi.; Jiang, J.; Liang, J.; Yu, D.; Cui, P.; Yang, H. Ferroelectrically controlled topological magnetic phase in a Janus-magnet-based multiferroic heterostructure. *Phys. Rev. Research* **2021**, 3, 043011.

(21) Wang, L.; Feng, Q.; Kim, Y.; Kim, R.; Lee, K. H.; Pollard, S. D.; Shin, Y. J.; Zhou, H.; Peng, W.; Lee, D.; Meng, W.; Yang, H.; Han, J. H.; Kim, M.; Lu, Q.; Noh, T. W. Ferroelectrically tunable magnetic skyrmions in ultrathin oxide heterostructures. *Nature Mater.* **2018**, 17, 1087–1094.

(22) Tsymbal, E. Y.; Panagopoulos, C. Whirling spins with a ferroelectric. *Nature Mater.* **2018**, 17, 1054–1055.

(23) Tikhonov, Yu.; Kondovych, S.; Mangeri, J.; Pavlenko, M.; Baudry, L.; Sené, A.; Galda, A.; Nakhmanson, S.; Heinonen, O.; Razumnaya, A.; Luk'yanchuk, I.; Vinokur, V. M. Controllable skyrmion chirality in ferroelectrics. *Scientific Reports* **2020**, 10, 8657.

(24) Wang, Y.; Sun, J.; Shimada, T.; Hirakata, H.; Kitamura, T.; Wang, J. Ferroelectric control of magnetic skyrmions in multiferroic heterostructures. *Phys. Rev. B* **2020**, 102, 014440.

(25) Gong, C.; Zhang, X. Two-dimensional magnetic crystals and emergent heterostructure devices. *Science* **2019**, 363, eaav4450.

(26) Gibertini, M.; Koperski, M.; Morpurgo, A. F.; Novoselov, K. S. Magnetic 2D materials and heterostructures. *Nat. Nanotechnol.* **2019**, 14, 408–419.

(27) Huang, P.; Zhang, P.; Xu, S.; Wang, H.; Zhang, X.; Zhang, H. Recent advances in two-dimensional ferromagnetism: materials synthesis, physical properties and device applications. *Nanoscale* **2020**, 12, 2309–2327.

(28) Jiang, X.; Liu, Q.; Xing, J.; Liu, N.; Guo, Y.; Liu, Z.; Zhao, J. Recent progress on 2D magnets: Fundamental mechanism, structural design and modification. *Appl. Phys. Rev.* **2021**, 8, 031305.

(29) Gong, C.; Li, L.; Li, Z.; Ji, H.; Stern, A.; Xia, Y.; Cao, T.; Bao, Wei; Wang, C.; Wang, Y.; Qiu, Z. Q.; Cava, R. J.; Louie, S. G.; Xia, J.; Zhang, X. Discovery of intrinsic ferromagnetism in two-dimensional van der Waals crystals. *Nature* **2017**, 546, 265–269.

(30) Wu, M. Two-dimensional van der Waals ferroelectrics: Scientific and technological opportunities. *ACS Nano* **2021**, 15, 9229–9237.

(31) Tsymbal, E. Y. Two-dimensional ferroelectricity by design. *Science* **2021**, 372, 1389–1390.

(32) Belianinov, A.; He, Q.; Dziaugys, A.; Maksymovych, P.; Eliseev, E.; Borisevich, A.; Morozovska, A.; Banys, J.; Vysochanskii, Y.; Kalinin, S. V. $CuInP_2S_6$ Room Temperature Layered Ferroelectric. *Nano Lett.* **2015**, 15, 3808–3814.

(33) Chang, K.; Liu, J.; Lin, H.; Wang, N.; Zhao, K.; Zhang, A.; Jin, F.; Zhong, Y.; Hu, X.; Duan, W.; Zhang, Q.; Fu, L.; Xue, Q.-K.; Chen, X.; Ji, S.-H. Discovery of robust in-plane ferroelectricity in atomic-thick SnTe. *Science* **2016**, 353, 274-278.

(34) Xiao, J.; Zhu, H.; Wang, Y.; Feng, W.; Hu, Y.; Dasgupta, A.; Han, Y.; Wang, Y.; Muller, D. A.; Martin, L. W.; Hu, P.; Zhang, X. Intrinsic two-dimensional ferroelectricity with dipole locking. *Phys. Rev. Lett.* **2018**, 120, 227601.

(35) Fei, Z.; Zhao, W.; Palomaki, T. A.; Sun, B.; Miller, M. K.; Zhao, Z.; Yan, J.; Xu, X.; Cobden D. H. Ferroelectric switching of a two-dimensional metal. *Nature* **2018**, 560, 336–339.

(36) Sharma, P.; Xiang, F.-X.; Shao, D.-F.; Zhang, D.; Tsymbal, E. Y.; Hamilton, A. R.; Seidel, J. A room-temperature ferroelectric semimetal. *Sci. Adv.* **2019**, 5, eaax5080.

(37) Yasuda, K.; Wang, X.; Watanabe, K.; Taniguchi, T.; Jarillo-Herrero, P. Stacking-engineered ferroelectricity in bilayer boron nitride. *Science* **2021**, 372, 1458–1462.

(38) Stern, M. V.; Waschitz, Y.; Cao, W.; Nevo, I.; Watanabe, K.; Taniguchi, T.; Sela, E.; Urbakh, M.; Hod, O.; Shalom, M. B. Interfacial ferroelectricity by van der Waals sliding. *Science* **2021**, 372, 1462–1466.

(39) Žutić, I.; Matos-Abiague, A.; Scharf, B.; Dery, H.; Belashchenko, K. Proximitized materials. *Mater. Today* **2019**, 22, 85-107.

(40) Park, T.-E.; Peng, L.; Liang, J.; Hallal, A.; Yasin, F. S.; Zhang, X.; Song, K. M.; Kim, S. J.; Kim, K.; Weig, M.; Schütz, G.; Finizio, S.; Raabe, J.;

(40) Yang, M.; Li, Q.; Chopdekar, R. V.; Stan, C.; Cabrini, S.; Choi, J. W.; Wang, S.; Wang, T.; Gao, N.; Scholl, A.; Tamura, N.; Hwang, C.; Wang, F.; Qiu, Z. Creation of skyrmions in van der Waals ferromagnet $Fe_3GeTe_2$ on $(Co/Pd)_n$ superlattice. *Sci. Adv.* **2020**, 6, eabb5157.

Garcia, K.; Xia, J.; Zhou, Y.; Ezawa, M.; Liu, X.; Chang, J.; Koo, H. C.; Kim, Y. D.; Chshiev, M.; Fert, A.; Yang, H.; Yu, X.; Woo, S. Néel-type skyrmions and their current-induced motion in van der Waals ferromagnet-based heterostructures. *Phys. Rev. B* **2021**, 103, 104410.

(41) Ding, B.; Li, Z.; Xu, G.; Li, Hang; Hou, Z.; Liu, E.; Xi, X.; Xu, F.; Yao, Y.; Wang, W. Observation of magnetic skyrmion bubbles in a van der Waals ferromagnet $Fe_3GeTe_2$. *Nano Lett.* **2020**, 2, 20.

(42) Wu, Y.; Zhang, S.; Zhang, J.; Wang, W.; Zhu, Y. L.; Hu, J.; Yin, G.; Wong, K.; Fang, C.; Wan, C.; Han, X.; Shao, Q.; Taniguchi, T.; Watanabe, K.; Zang, J.; Mao, Z.; Zhang, X.; Wang, K. L. Néel-type skyrmion in $WTe_2/Fe_3GeTe_2$ van der Waals heterostructure. *Nat. Commun.* **2020**, 11, 3860.

(43) Yang, M.; Li, Q.; Chopdekar, R. V.; Dhall, R.; Turner, J.; Carlström, J. D.; Ophus, C.; Klewe, C.; Shafer, P.; N'Diaye, A. T.; Choi, J. W.; Chen, G.; Wu, Y. Z.; Hwang, C.; Wang, F.; Qiu, Z. Q. Creation of skyrmions in van der Waals ferromagnet $Fe_3GeTe_2$ on $(Co/Pd)_n$ superlattice. *Sci. Adv.* **2020**, 6, eabb5157.

(44) Osada, M.; Sasaki, T. The rise of 2D dielectrics/ferroelectrics. *APL Materials* **2019**, 7, 120902.

(45) Guan, Z.; Hu, H.; Shen, X.; Xiang, P.; Zhong, N.; Chu, J.; Duan, C. Recent progress in two-dimensional ferroelectric materials. *Adv. Electron. Mater.* **2020**, 6, 1900818.

(46) Cui, C.; Hu, W.-J.; Yan, X.; Addiego, C.; Gao, W.; Wang, Y.; Wang, Z.; Li, L.; Cheng, Y.; Li, P.; Zhang, X.; Alshareef, H. N.; Wu, T.; Zhu, W.; Pan, X.; Li, L.-J. Intercorrelated in-plane and out-of-plane ferroelectricity in ultrathin two-dimensional layered semiconductor $In_2Se_3$. *Nano Lett.* **2018**, 18, 1253–1258.

(47) Zhou, Y.; Wu, D.; Zhu, Y.; Cho, Y.; He, Q.; Yang, X.; Herrera, K.; Chu, Z.; Han, Y.; Downer, M. C.; Peng, H.; Lai, K. Out-of-plane piezoelectricity and ferroelectricity in layered α-$In_2Se_3$ nanoflakes. *Nano Lett.* **2017**, 17, 5508–5513.

(48) Xue, F.; Hu, W.; Lee, K.-C.; Lu, L.-S.; Zhang, J.; Tang, H.-L.; Han, A.; Hsu, W.-T.; Tu, S.; Chang, W.-H.; Lien, C.-H.; He, J.-H.; Zhang, Z.; Li, L.-J.; Zhang, X. Room-temperature ferroelectricity in hexagonally layered α-$In_2Se_3$ nanoflakes down to the monolayer limit. *Adv. Funct. Mater.* **2018**, 28, 1803738.

(49) Ding, W.; Zhu, J.; Wang, Z.; Gao, Y.; Xiao, D.; Gu, Y.; Zhang, Z.; Zhu, W. Prediction of intrinsic two-dimensional ferroelectrics in $In_2Se_3$ and other $III_2-VI_3$ van der Waals materials. *Nat. Commun.* **2017**, 8, 14956.

(50) Gong, C.; Kim, E. M.; Wang, Y.; Lee, G.; Zhang, X. Multiferroicity in atomic van der Waals heterostructures. *Nat. Commun.* **2019**, 10, 2657.

(51) Wang, Z.; Zhu, W. Tunable band alignments in 2D ferroelectric α-$In_2Se_3$ based Van der Waals heterostructures. *ACS Appl. Electron. Mater.* **2021**, 3, 5114–5123.

(52) Wan, S.; Li, Y.; Li, W.; Mao, X.; Wang, C.; Chen, C.; Dong, J.; Nie, A.; Xiang, J.; Liu, Z.; Zhu, W.; Zeng, H. Nonvolatile ferroelectric memory effect in ultrathin α-$In_2Se_3$. *Adv. Funct. Mater.* **2019**, 29, 1808606.

(53) Li, Y.; Chen, C.; Li, W.; Mao, X.; Liu, H.; Xiang, J.; Nie, A.; Liu, Z.; Zhu, W.; Zeng, H. Orthogonal electric control of the out-of-plane field-effect in 2D ferroelectric α-$In_2Se_3$. *Adv. Electron. Mater.* **2020**, 6, 2000061.

(54) Shao, D.-F.; Ding, J.; Gurung, G.; Zhang, S.-H.; Tsymbal, E. Y. Interfacial crystal Hall effect reversible by ferroelectric polarization. *Phys. Rev. Applied* **2021**, 15, 024057.

(55) Yang, H.; Boulle, O.; Cros, V.; Fert, A.; Chshiev, M. Controlling Dzyaloshinskii-Moriya interaction via chirality dependent atomic-layer stacking, insulator capping and electric field. *Scientific Reports* **2018**, 8, 2356.

(56) Chen, D.; Sun, W.; Li, H.; Wang, J.; Wang, Y. Tunable magnetic anisotropy and Dzyaloshinskii-Moriya interaction in an ultrathin van der Waals $Fe_3GeTe_2/In_2Se_3$ heterostructure. *Front. Phys.* **2020**, 8, 402.

(57) Li, C.; Yao, X.; Chen, G. Writing and deleting skyrmions with electric fields in a multiferroic heterostructure. *Phys. Rev. Research* **2021**, 3, L012026.

(58) Deng, Y.; Yu, Y.; Song, Y.; Zhang, J.; Wang, N. Z.; Sun, Z.; Yi, Y.; Wu, Y. Z.; Wu, S.; Zhu, J.; Wang, J.; Chen, X. H.; Zhang, Y. Gate-tunable room-temperature ferromagnetism in two-dimensional $Fe_3GeTe_2$. *Nature* **2018**, 563, 94–99.

(59) Fei, Z.; Huang, B.; Malinowski, P.; Wang, W.; Song, T.; Sanchez, J.; Yao, W.; Xiao, D.; Zhu, X.; May, A. F.; Wu, W.; Cobden, D. H.; Chu, J.-H.; Xu, X. Two-dimensional itinerant ferromagnetism in atomically thin $Fe_3GeTe_2$. *Nature Mater.* **2018**, 17, 778–782.

(60) Chen, B.; Yang, J.; Wang, H.; Imai, M.; Ohta, H.; Michioka, C.; Yoshimura, K.; Fang, M. Magnetic properties of layered itinerant electron ferromagnet $Fe_3GeTe_2$. *J. Phys. Soc. Jpn.* **2013**, 82, 124711.

(61) Zhuang, H. L.; Kent, P. R. C.; Hennig, R. G. Strong anisotropy and magnetostriction in the two-dimensional Stoner ferromagnet $Fe_3GeTe_2$. *Phys. Rev. B* **2016**, 93, 134407.

(62) Laref, S.; Kim, K.; Manchon, A. Elusive Dzyaloshinskii-Moriya interaction in monolayer Fe3GeTe2. *Phys. Rev. B* **2020**, 102, 060402(R).

(63) Yang, H.; Thiaville, A.; Rohart, S.; Fert, A.; Chshiev, M. Anatomy of Dzyaloshinskii-Moriya interaction at Co/Pt interfaces. *Phys. Rev. Lett.* **2015**, 115, 267210.

(64) Tao, X.; Gu, Y. Crystalline–crystalline phase transformation in two-dimensional $In_2Se_3$ thin layers. *Nano Lett.* **2013**, 13, 3501–3505.

(65) Wang, X.S.; Yuan, H.Y.; Wang, X.R. A theory on skyrmion size. *Commun. Phys.* **2018**, 1, 31.

(66) Wanga, Y.-P.; Chen, X.-Y.; Long, M.-Q. Modifications of magnetic anisotropy of $Fe_3GeTe_2$ by the electric field effect. *Appl. Phys. Lett.* **2020**, 116, 092404.

(67) Park, S. Y.; Kim, D. S.; Liu, Y.; Hwang, J.; Kim, Y.; Kim, W.; Kim, J.-Y.; Petrovic, C.; Hwang, C.; Mo, S.-K.; Kim, H.-J.; Min, B.-C.; Koo, H. C.; Chang, J.; Jang, C.; Choi, J. W.; Ryu, H. Controlling the magnetic anisotropy of the van der waals ferromagnet $Fe_3GeTe_2$ through hole doping. *Nano Lett.* **2020**, 20, 95–100.

(68) Tan, C.; Lee, J.; Jung, S.-G.; Park, T.; Albarakati, S.; Partridge, J.; Field, M. R.; McCulloch, D. G.; Wang, L.; Lee, C. Hard magnetic properties in nanoflake van der Waals $Fe_3GeTe_2$. *Nat. Commun.* **2018**, 9, 1554.

(69) Srivastava, T.; Schott, M.; Juge, R.; Křižáková, V.; Belmeguenai, M.; Roussigné, Y.; Bern-Mantel, A.; Ranno, L.; Pizzini, S.; Chérif, S.-M.; Stashkevich, A.; Auffret, S.; Boulle, O.; Gaudin, G.; Chshiev, M.; Baraduc, C.; Béa, H. Large-Voltage Tuning of Dzyaloshinskii–Moriya Interactions: A Route toward Dynamic Control of Skyrmion Chirality. *Nano Lett.* **2018**, 18, 4871–4877.